# A CHARACTERIZATION OF GEODETIC GRAPHS IN TERMS OF THEIR EMBEDDED EVEN GRAPHS

**Frasser C.E**.
*Ph.D. in Engineering Sciences, Odessa National Polytechnic University, Ukraine*

**Abstract.**
The problem of finding the general classification of geodetic graphs is still open. We believe that one of the obstacles to attain this goal is that geodetic graphs lack a structural description. In other words, their fundamental properties have not yet been established in terms of the description of the complete graphs, paths and cycles contained within them. The absence of this information makes their generation and enumeration (as inherent parts of their general classification) a difficult task. This paper examines the structural qualities of geodetic graphs using their so-called embedded even graphs. To this effect, the necessary and sufficient conditions for eliminating the nongeodecity of each pair of $C$-opposite vertices in an even cycle $C$ have been formulated, while the bigeodecity of the embedded even graphs of a geodetic graph has been established. In a sense, this allows us to arrive at the conclusion that the basic building blocks of geodetic graphs are precisely this class of bigeodetic ones.

**Introduction**

When beginning the design process of a computer network, the total set of information sources, the set of users, and the pairwise distances between them are generally known. Information sources include complex knowledge bases, encompassing databases and specialized software, which store enormous amounts of information. This information is accessed using specific computerized methods. Information users may include specific organizations and entities that possess the necessary technical means to connect to the network. The information obtained, after processing and analysis, is organized into a complex knowledge base. Interaction between these knowledge bases is established through transmission channels, which must be organized to maximize the information transmission speed. Clearly, this condition is met only when each pair of network centers is connected by a transmission channel. If we assign to each information center a vertex of an undirected graph, whose edges correspond to the transmission channels, then we obtain a complete graph, which, in fact, is a geodetic one. Using this connection arrangement between information centers minimizes the time required for information transmission, which is equivalent to maximizing information transmission speed. However, as the number of centers increases, the material costs for creating the network also increase, and in this case, the structure of a complete graph is no longer convenient [3]. It is evident that the construction of a geodetic network of the smallest possible diameter will be preferable. This network contains embedded even graphs with a special structure whose description we study next as an attempt to characterize geodetic graphs from a different perspective. We think, this could contribute to have a better understanding of how to solve the open problem of finding the general classification of geodetic graphs [2].

# On some Structural Properties of Geodetic Graphs

In an even cycle, any pair of diametrically opposite vertices are joined by two geodesics determined by the two paths of equal length that make up the cycle. From a network perspective, an even cycle leads to the duplication of information transmission channels. If a high level of network reliability is required [3], then such a structure may be the most suitable. However, if minimizing network setup costs while maintaining a high data transmission speed is necessary, then such a structure is not appropriate. It is possible to eliminate the duplication of transmission channels by introducing a system of chords into the even cycle, whose result is what we call an embedded even graph. In order to describe the fundamental properties of embedded even graphs in a geodetic graph, we need to introduce some fundamental concepts.

The set of *vertices* and *edges* of a graph *G* are denoted by *V(G)* and *E(G)*, respectively. A *path* from $v_0$ to $v_n$ is a sequence $v_0 v_1 \ldots v_n$ of different vertices of *G* and is denoted by $P(v_0, v_n)$. The *number* of edges of a path *P* in *G* determines the *length* of this path and is represented by $|P|$. A graph *G* is *connected* if there exists a path in *G* between any pair of its vertices. The *distance* between $u, v \in V(G)$, denoted by $d_G(u, v)$, is the length of the shortest path connecting vertices $u, v$. The *diameter* of *G*, denoted $d(G)$, is the greatest distance between any pair of its vertices. The number of edges incident to a vertex *v* of *G* is called the *degree* of *v* and is denoted by $\deg(v)$. In a graph *G*, any vertex *v* having $\deg(v) \geq 3$ is called a *node*. A *segment* is a path whose only nodes are its end vertices *u* and *v*. Two graphs $G_1$ and $G_2$ are said to be *homeomorphic* if and only if there exists a graph which can be obtained by insertion of vertices onto the edges of one or both of $G_1$ and $G_2$. Note that any connected graph is homeomorphic to itself. A *complete graph* is a graph in which every pair of vertices are joined by an edge. The complete graph with *n* vertices is denoted by $K_n$. A *cycle* is a sequence $v_0 v_1 \ldots v_n v_0$ of different vertices of *G*. A cycle *C* of *G* is called *even* (*odd*) if its length |C| is even (odd). Assume that *C* is even. If $u, v \in V(C)$, $u \neq v$, then there exist exactly two paths belonging to *C* joining vertices *u* and *v*. One of these paths is denoted by $C(u, v)$, and the other, $C(v, u)$. Two vertices $x, y$ of an even cycle *C* are called *C-opposite* if $|C(x, y)| = |C(y, x)|$. A path $A(u, v)$ joining no more than two vertices $u, v$ of a cycle *C* is called a *chord* if $|A(u, v)| < |C(u, v)|$ and $|A(u, v)| < |C(v, u)|$. A *subgraph H* of a graph *G* is a graph whose vertices and edges are subsets of those of *G*. The shortest path $\Gamma(u, v)$ joining the vertices $u, v$ of *G* is called a *geodesic*. A graph *G* is *geodetic* if there exists a unique geodesic between every pair of its vertices. A graph *G* is called *K-geodetic* if between each pair of its vertices there are, at most, *K* distinct geodesics. A *K*-geodetic graph is geodetic if $K = 1$, *bigeodetic* if $K = 2$ and *trigeodetic* if $K = 3$.

**Lemma 1.** ([5]) *A connected graph G is geodetic if and only if G does not contain an even cycle C such that for every pair of C-opposite vertices $u, v$ $d_G(u, v) = d(C)$.* □

**Theorem 1.** ([4]) *Let G be homeomorphic to the complete graph $K_4$; G is geodetic if and only if:*
  1. *The six segments of G are geodesics.*
  2. *Each cycle of G that contains exactly three segments is odd.*
  3. *All cycles of G that contain exactly four segments have equal length.* □

Let *H* be the graph obtained as a result of the union of an even cycle *C* that does not cut itself and a set of *n* vertex-disjoint chords $A_1 = A(x_1, x_{n+1})$, $A_2 = A(x_2, x_{n+2}), \ldots, A_n = A(x_n, x_{2n})$, whose end vertices correspond to the pairs $(x_1, x_{n+1}), (x_2, x_{n+2}), \ldots, (x_n, x_{2n})$, where $x_1, x_2, \ldots, x_n, x_{n+1}, x_{n+2}, \ldots, x_{2n}$ is a successive cyclic labeling of the degree-three vertices (nodes) of $H$ in the clockwise direction, which begins at some arbitrarily chosen node. We then define an *arc* $C(u, v)$ in $H$ as the path formed by the vertices and edges of the even cycle $C$ joining two nodes $u$ and $v$ of $H$. We denote by $|C(x, y)|$ the length of the shortest path in $C$ joining any two arbitrary vertices $x$ and $y$ of $C$. This is also valid in the case where $C(x, y)$ is an arc. Two chords $A$ and $B$ of $H$ are said to be *adjacent* if, in the arcs of the even cycle $C$ subtended by the end vertices of these chords, there are no end vertices of other chords. The cycles generated by a pair of adjacent chords and their corresponding subtended arcs have a length that we will designate as follows:

$$|C_i^j| = |C(x_i, x_j)| + |A(x_i, x_{n+i})| + |A(x_j, x_{n+j})| + |C(x_{n+i}, x_{n+j})|,$$

for $i = 1, \ldots, n$, where $x_i, x_j$ are two consecutive nodes of $H$, with $j > i$ for $i \leq n - 1$ and $j = 1$ for $i = n$.

An embedded even graph $H$ of a geodetic graph $G$ containing a minimal even cycle $C$ is a subgraph $H$ of $G$ obtained by combining the even cycle $C$ and its chord system (in the sense explained above) such that in $H$ no pair of adjacent chords and the subtended arcs of $C$, or chord and arc, form an even cycle of length less than $|C|$.

Based on the concepts previously described, our main theorem follows.

**Theorem 2.** *Let $H$ be an embedded even graph that has a system of $n$ vertex-disjoint chords and whose even cycle $C$ has length $2L$ for any integer $L \geq 2$, where $2 \leq n \leq L$. The chord system of $H$ eliminates the nongeodecity of all pairs of C-opposite vertices and preserves the geodecity of all pairs of non-C-opposite vertices if and only if the following conditions are met in $H$:*

1. *Every cycle consisting of an arc and a chord has odd length,*

2. $|C_1^2| = |C_2^3| = |C_3^4| = \cdots = |C_{n-1}^n| = |C_n^1| = |C| = 2L$

*Proof*: *Necessity* will be carried out by contradiction.

*Case 1*. Suppose there exists an embedded even graph $H$, whose chord system eliminates the nongeodecity of all pairs of *C*-opposite vertices and preserves the geodecity of all pairs of non-*C*-opposite vertices, in which condition 2 satisfies, but in which there exists at least one even cycle $C^*$ formed by the arc $C(x_i, x_{n+i})$ or the arc $C(x_{n+i}, x_i)$ and a chord $A_i$, for some $i = 1, \ldots, n$. Then $C^*$ in $H$ is an even cycle of length less than $2L$ formed by an arc of cycle $C$ and a chord. But this contradicts the definition of $H$ being an embedded even graph.

*Case 2*. Let us now suppose that there exists an embedded even graph $H$, whose chord system eliminates the non-geodecity of all pairs of *C*-opposite vertices and preserves the geodecity of all pairs of non*C*-opposite vertices, in which condition 1 satisfies, but in which there exist even cycles $C_i^j$ such that $|C_i^j| \neq 2L$. If for some pair $i, j$, $|C_i^j| < 2L$, then $H$ contains a pair of adjacent chords with their respective subtended arcs, which form an even cycle of length less than $|C|$. This

contradicts the definition of $H$. Therefore, suppose that $|C_i^j| > 2L$. Let us examine one of the cycles $C_i^j$. Without loss of generality, suppose that $C_i^j$ is the cycle $C_1^2$ (Fig. 1).

Let us select in $H$ the vertex $v \in C$, $v \in C_1^2$ that lies on the odd cycle formed by the arc $C(x_1, x_{n+1})$ and the chord $A_1 = A(x_1, x_{n+1})$ such that

$$|C(v, x_{n+1})| = |A_1| + |C(x_1, v)| - 1 \qquad (1)$$

Select now vertex $u$ so that $u$ and $v$ are $C_1^2$-opposite vertices and, at the same time, are $C$-opposite vertices in $H$. Then

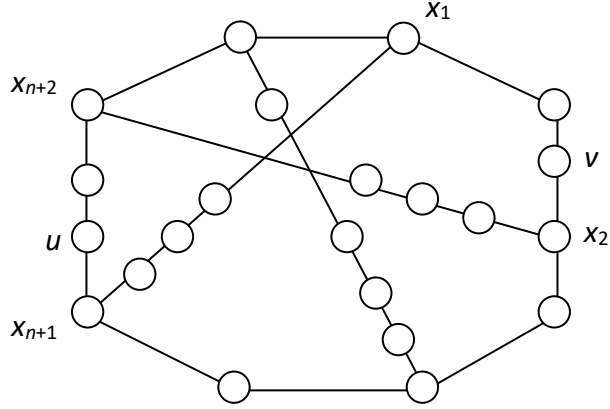

Fig. 1.

$$|C(u, x_{n+1})| + |C(x_{n+1}, v)| = L,$$

that is,

$$|C(x_{n+1}, v)| = L - |C(u, x_{n+1})| \qquad (2)$$

Equating (1) to (2), we obtain

$$|A_1| + |C(x_1, v)| - 1 = L - |C(u, x_{n+1})|,$$

which means that

$$|C(u, x_{n+1}) + |A_1| + |C(x_1, v)| = L + 1. \qquad (3)$$

Since $|C_1^2| > 2L$, that is, $|C_1^2| \geq 2L + 1$ and $|C(u, x_{n+1}) + |A_1| + |C(x_1, v)| = L + 1$,

$$|C(v, x_2)| + |A_2| + |C(x_{n+2}, u)| \geq L. \qquad (4)$$

Knowing that in $H$, $|\Gamma(u, v)|$ has to be less than or equal to $L - 1$, correlations (3) and (4) imply that the only geodesic $\Gamma(u, v)$ does not lie on $C_1^2$. But it does not lie on $C$ either because $u$ and $v$ are $C$-opposite vertices. Considered now the cycle $C^*$ formed by the chord $A_1$, one of the chords $A_j, j = n, n - 1, \ldots, 3$ and the arcs subtended by them. Note that $|C^*| \geq 2L$. It follows that for the only geodesic $\Gamma(u, v)$ (Fig. 1)

$$|\Gamma(u, v)| = |C^*| - |A_1| + |C(x_1, v)| + |C(u, x_{n+1})| \leq L - 1,$$

which is equivalent to

$$|C^*| \leq |A_1| - |C(x_1, v)| - |C(u, x_{n+1})| + L - 1$$

$$|C^*| \leq 2|A_1| - (|A_1| + |C(x_1, v)| + |C(u, x_{n+1})|) + L - 1. \tag{5}$$

Substituting (3) into (5), we obtain

$$|C^*| \leq 2|A_1| - L - 1 + L - 1$$

$$2L \leq |C^*| \leq 2|A_1| - 2$$

$$L \leq |A_1| - 1 < |A_1|$$

$$L < |A_1|,$$

which means that the length of a chord is greater than the length of any of the two corresponding arcs on $C$. This contradicts the above definition of a chord.

Lastly, consider the cycle $C^{**}$ formed by the chord $A_2$ one of the chords $A_j, j = 3, 4, \ldots, n$ and the arcs subtended by them. In this case,

$$|\Gamma(u, v)| = |C^{**}| - |A_2| + |C(v, x_2)| + |C(x_{n+2}, u)| \leq L - 1$$

$$|C^{**}| \leq |A_2| - |C(v, x_2)| - |C(x_{n+2}, u)| + L - 1.$$

$$|C^{**}| \leq 2|A_2| - (|A_2| + |C(v, x_2)| + |C(x_{n+2}, u)|) + L - 1 \tag{6}$$

From (4) and (6), we now obtain

$$2L \leq |C^{**}| \leq 2|A_2| - (|A_2| + |C(v, x_2)| + |C(x_{n+2}, u)|) + L - 1 \leq 2|A_2| - L + L - 1$$

$$2L \leq 2|A_2| - 1 < 2|A_2|$$

$$L < |A_2|,$$

which is again a contradiction.

From all of the above, we conclude that the geodesic $\Gamma(u, v)$ lies either on $C$, $C_1^2$, or both.

Since according to (4) $|C(v, x_2)| + |A_2| + |C(x_{n+2}, u)| \geq L$,

either $|C(v, x_2)| + |A_2| + |C(x_{n+2}, u)| = L$ or $|C(v, x_2)| + |A_2| + |C(x_{n+2}, u)| > L$.

If $|C(v, x_2)| + |A_2| + |C(x_{n+2}, u)| = L$, then $H$ contains three geodesics $\Gamma(u, v)$ of length $L$: one is contained in $C_1^2$ and the other two in $C$. This contradicts our assumption.

If $|C(v, x_2)| + |A_2| + |C(x_{n+2}, u)| > L$, then $H$ contains two geodesics $\Gamma(u, v)$ of length $L$, both contained in $C$, which again contradicts our assumption.

No contradictions arise only in the case when

$$|C(v, x_2)| + |A_2| + |C(x_{n+2}, u)| = L - 1$$

In this case, the pair of adjacent chords $A_1, A_2$ and their subtended arcs form a cycle in which $H$ has a unique geodesic $\Gamma(u, v)$ of length $L - 1$ and

$$\begin{aligned} C_1^2 &= (|C(u, x_{n+1}) + |A_1| + |C(x_1, v)|) + (|C(v, x_2)| + |A_2| + |C(x_{n+2}, u)|) \\ &= (L + 1) + (L - 1) = 2L. \end{aligned}$$

The same reasoning can be extended to any cycle $C_i^j$. So, we conclude that

$$|C_1^2| = |C_2^3| = |C_3^4| = \cdots = |C_{n-1}^n| = |C_n^1| = |C| = 2L$$

Let us now prove *sufficiency*.

Assume that $H$ satisfies conditions 1 and 2 of Theorem 2.

*Case 1*. If two $C$-opposite vertices $u$ and $v$ of $H$ are the end vertices of the chord $A_j$, for a given $j$, then $d_H(u, v) = |A_j| < L$ and the chord $A_j$ is the only geodesic in $H$.

*Case 2*. Let $u, v$ be $C$-opposite vertices that are not, at the same time, the end vertices of the chord $A_j$, where $u, v \in C_j^{j+1}$. If $u$ and $v$ are not $C_j^{j+1}$-opposite vertices, then, taking into account condition 2, that is, $C_j^{j+1} = 2L$, we conclude that $d_H(u, v) < L$ and the only geodesic $\Gamma(u, v)$ lies on $C_j^{j+1}$. If $u$ and $v$ are $C_j^{j+1}$-opposite vertices (Fig. 2), then let $C_j^{j+1} = C_1^2$. We then obtain

$$|C(v, x_1)| + |A_1| + |C(x_{n+1}, u)| = |C(u, x_{n+2})| + |A_2| + |C(x_2, v)| = L;$$

this implies that

$$|C(v, x_1)| + |A_1| = L - |C(x_{n+1}, u)|. \qquad (7)$$

At the same time,

$$|C(v, x_{n+1})| + |C(x_{n+1}, u)| = |C(u, x_1)| + |C(x_1, v)| = L;$$

this implies that

$$|C(v, x_{n+1})| = L - |C(x_{n+1}, u)|. \tag{8}$$

Equating (7) to (8), we now get

$$|C(v, x_{n+1})| = |C(v, x_1)| + |A_1|,$$

that is, the chord $A_1$ and the arc $C(x_1, x_{n+1})$ form an even cycle of length less than $|C| = 2L$, which contradicts condition 1.

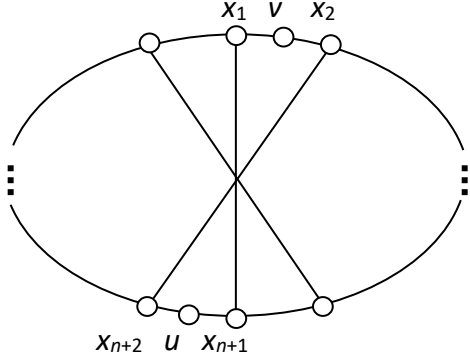

Fig. 2.

*Case 3.* If $u$, $v$ are $C$-opposite vertices in $G$ that do not belong to any of the cycles $C_1^2, C_2^3, \ldots, C_{n-1}^n, C_n^1$, then both $u$ and $v$ belong to one of the cycles formed by an arc of the cycle $C$ and a chord. Since this odd cycle has a length less than $2L$, then $d_H(u, v) < L$. Consequently, for the even cycle $C$ and any pair of its $C$-opposite vertices $u$ and $v$, we obtained that $d_H(u, v) < d(C)$ and the proof follows immediately from Lemma 1. □

**Corollary 1.** *If the embedded even graph H of Theorem 2 has $n = 2$ chords, then H is geodetic.*

*Proof.* Clearly, when $n = 2$, $H$ is homeomorphic to the complete graph $K_4$.

i) $C_1^2, C_2^1, C$ all contain exactly four segments. Condition 2 of Theorem 2 guarantees that $|C_1^2| = |C_2^1| = |C|$, that is, the three cycles containing four segments have equal length and condition 3 of Theorem 1 follows.
ii) The cycles obtained as a result of joining a chord an its two corresponding arcs each contains exactly three segments. Condition 1 of Theorem 2 guarantees that all of them are odd. So, condition 2 of Theorem 1 also follows.
iii) Statements i) and ii) imply, by Theorem 2, that the four segments lying on $C$ are all geodesics. The other 2 segments are the chords $A_1$ and $A_2$. But $|A_1| < L$ and $|A_2| < L$. This means that both of them are geodesics in $H$. Consequently, the six segments of $G$ are geodesics. Condition 1 of Theorem 1 follows as well.

Therefore, $H$ satisfies conditions 1, 2 and 3 of Theorem 1. Corollary 1 now follows. □

**Corollary 2.** *If in H the number of chords n is such that $3 \leq n \leq L$, then H is bigeodetic.*

*Proof.* By Theorem 2, every pair of vertices belonging to $C$ has a unique geodesic in $H$. Let $u, v$ be two vertices belonging to the chord system of $H$ such that $u \in A_i$, $v \in A_j$, $i \neq j$. Knowing that the pair of chords $A_i$ and $A_j$ (including chords $A_i$ and $A_j$ that are adjacent) lie on two cycles $C'$ and $C''$ formed by these two chords and their two corresponding inner and outer arcs subtended by them, it can be verified that if $u$ and $v$ are not $C'$-opposite, then the pair of vertices $u$ and $v$ are joined, either in $C'$ by a unique geodesic $\Gamma(u,v)$ or in $C''$ by a unique geodesic $\Gamma(u,v)$ (if that geodesic lies on $C''$). If vertices $u$ and $v$ are $C'$-opposite, then $u, v$ are joined, either in $C'$ by two geodesics $\Gamma(u,v)$ or in $C''$ by a unique geodesic $\Gamma(u,v)$ (if that geodesic lies on $C''$). Therefore, $H$ containing $n$ chords, $3 \leq n \leq L$, is bigeodetic. □

**Corollary 3.** *The embedded even cycle H of Theorem 2 is either geodetic if $n = 2$ or bigeodetic if $3 \leq n \leq L$.* □

**Theorem 3.** *Any minimal even cycle C (an even cycle C of minimal length) contained in a geodetic graph G generates an embedded even graph H (in the sense of Theorem 2) that is a subgraph of G.*

*Proof.* Since $C$ is a minimal even cycle, there are no even cycles of length less than $|C|$ in $G$. Now, since $G$ is geodetic, by Lemma 1, it must happen that $d_G(x,y) < d(C)$ for any pair of $C$-opposite vertices $x, y$ in $G$. Therefore, the vertices of $C$ must be joined in $G$ with a set of paths that eliminates the nongeodecity of all pairs of $C$-opposite vertices. According to Theorem 2, in $G$, a system of chords among those paths must be found for $C$ satisfying conditions 1 and 2. Therefore, for $C$ as a minimal even cycle contained in $G$, graph $G$ must generate an embedded even graph $H$ as its subgraph. □

**Corollary 4.** *Let G be a connected graph. If in G there exists at least one minimal even cycle C that does not generate the embedded even graph H of Theorem 2 as a subgraph of G, then G is not geodetic.* □

## Concluding Remarks

So far, a classification of geodetic graphs includes families of the following type: odd cycle graphs, cacti lacking even cycles, complete graphs $K_n$ and the homeomorphic to them $K_n^i$, geodetic graphs involving section subgraphs of complete graphs (among them, barbell graphs, lollypop graphs, windmill graphs, triangular snake graphs), both the Petersen graph and Hoffman-Singleton graph together with some special families homeomorphic to them, trees, clique trees, some special types of graphs with cliques of diameter two constructed from a generalization of affine planes, some graphs of diameter three having a similar structural description to that of a Moore graph, graphs of diameter four and five constructed using triple systems and symmetric block designs and, probably, other families that are out there, but we do not know of. Generally speaking, the classification not only of geodetic graphs, but also of bigeodetic and trigeodetic graphs are unsolved problems in graph theory [1]. Every geodetic graph is also bigeodetic. With regard to the less known bigeodetic graphs, specifically the ones that are not geodetic, only few families are known: the ones described in this article named embedded even graphs, separable bigeodetic graphs, perfect bigeodetic graphs, bigeodetic graphs contractible to a geodetic one, the ladder-like structures, and bigeodetic graphs constructed using triple systems and symmetric block designs. Regarding trigeodetic graphs, they have been much less studied, but two of their constructions are described in [1]. Finally, $K$-geodetic graphs have several important applications. In network topology design when controlled path redundancy make them suitable if path diversity is taking into account, in routing algorithms when the property of geodecity makes easier to take routing decisions by controlling the number of shortest paths to be considered, in fault-tolerant systems if the $K$-geodetic network possesses a higher index of connectivity to ensure network resilience, while the controlled number of shortest paths makes easier rerouting after network failures, and in distributed computing systems for an easy design of network communication that require predictable patterns [3].

## References

1. C.E. Frasser, Geodetic Graphs and Block Designs: A Survey, Journal of Science. Lyon 62 (2025) pp. 20-24.
2. C.E. Frasser, The Open Problem of Finding a General Classification of Geodetic Graphs, German International Journal of Modern Science 36 (2022) pp. 7-9.
3. C.E. Frasser, Topological Analysis and Synthesis related to Certain Classes of $K$-Geodetic Computer Networks, German International Journal of Modern Science 85 (2024) pp. 29-36.
4. J.G. Stemple, Geodetic Graphs Homeomorphic to a Complete Graph, Annals of New York Academy of Sciences 319 (1979) pp. 512-517.
5. J.G. Stemple, M.E. Watkins, On Planar Geodetic Graphs, J. Combin. Theory 4 (1968) pp. 101-117.